
%
%
%
%
\message
{JNL.TEX version 0.9 as of 3/27/86.  Report bugs and problems to Doug Eardley.}



\font\twelverm=cmr10 scaled 1200    \font\twelvei=cmmi10 scaled 1200
\font\twelvesy=cmsy10 scaled 1200   \font\twelveex=cmex10 scaled 1200
\font\twelvebf=cmbx10 scaled 1200   \font\twelvesl=cmsl10 scaled 1200
\font\twelvett=cmtt10 scaled 1200   \font\twelveit=cmti10 scaled 1200

\skewchar\twelvei='177   \skewchar\twelvesy='60


\def\twelvepoint{\normalbaselineskip=12.4pt plus 0.1pt minus 0.1pt
  \abovedisplayskip 12.4pt plus 3pt minus 9pt
  \belowdisplayskip 12.4pt plus 3pt minus 9pt
  \abovedisplayshortskip 0pt plus 3pt
  \belowdisplayshortskip 7.2pt plus 3pt minus 4pt
  \smallskipamount=3.6pt plus1.2pt minus1.2pt
  \medskipamount=7.2pt plus2.4pt minus2.4pt
  \bigskipamount=14.4pt plus4.8pt minus4.8pt
  \def\rm{\fam0\twelverm}          \def\it{\fam\itfam\twelveit}%
  \def\sl{\fam\slfam\twelvesl}     \def\bf{\fam\bffam\twelvebf}%
  \def\mit{\fam 1}                 \def\cal{\fam 2}%
  \def\tt{\twelvett}
  \textfont0=\twelverm   \scriptfont0=\tenrm   \scriptscriptfont0=\sevenrm
  \textfont1=\twelvei    \scriptfont1=\teni    \scriptscriptfont1=\seveni
  \textfont2=\twelvesy   \scriptfont2=\tensy   \scriptscriptfont2=\sevensy
  \textfont3=\twelveex   \scriptfont3=\twelveex  \scriptscriptfont3=\twelveex
  \textfont\itfam=\twelveit
  \textfont\slfam=\twelvesl
  \textfont\bffam=\twelvebf \scriptfont\bffam=\tenbf
  \scriptscriptfont\bffam=\sevenbf
  \normalbaselines\rm}



\def\beginlinemode{\endmode
  \begingroup\parskip=0pt \obeylines\def\\{\par}\def\endmode{\par\endgroup}}
\def\beginparmode{\endmode
  \begingroup \def\endmode{\par\endgroup}}
\let\endmode=\par
{\obeylines\gdef\
{}}
\def\singlespace{\baselineskip=\normalbaselineskip}

\def\oneandahalfspace{\baselineskip=\normalbaselineskip
  \multiply\baselineskip by 3 \divide\baselineskip by 2}
\def\doublespace{\baselineskip=\normalbaselineskip \multiply\baselineskip by 2}

\newcount\firstpageno
\firstpageno=2
\footline=
{\ifnum\pageno<\firstpageno{\hfil}\else{\hfil\twelverm\folio\hfil}\fi}
\def\toppageno{\global\footline={\hfil}\global\headline
  ={\ifnum\pageno<\firstpageno{\hfil}\else{\hfil\twelverm\folio\hfil}\fi}}
\let\rawfootnote=\footnote		
\def\footnote#1#2{{\rm\singlespace\parindent=0pt\parskip=0pt
  \rawfootnote{#1}{#2\hfill\vrule height 0pt depth 6pt width 0pt}}}
\def\raggedcenter{\leftskip=4em plus 12em \rightskip=\leftskip
  \parindent=0pt \parfillskip=0pt \spaceskip=.3333em \xspaceskip=.5em
  \pretolerance=9999 \tolerance=9999
  \hyphenpenalty=9999 \exhyphenpenalty=9999 }
\def\dateline{\rightline{\ifcase\month\or
  January\or February\or March\or April\or May\or June\or
  July\or August\or September\or October\or November\or December\fi
  \space\number\year}}
\def\today{\ifcase\month\or
  January\or February\or March\or April\or May\or June\or
  July\or August\or September\or October\or November\or December\fi
  \space\number\day, \number\year}
\def\received{\vskip 3pt plus 0.2fill
 \centerline{\sl (Received\space\ifcase\month\or
  January\or February\or March\or April\or May\or June\or
  July\or August\or September\or October\or November\or December\fi
  \qquad, \number\year)}}


\hsize=6.5truein
\hoffset=0truein
\vsize=8.9truein
\voffset=0truein
\parskip=\medskipamount
\def\\{\cr}
\twelvepoint		
\doublespace		
\overfullrule=0pt	




\def\title			
  {\null\vskip 3pt plus 0.2fill
   \beginlinemode \doublespace \raggedcenter \bf}

\def\author			
  {\vskip 3pt plus 0.2fill \beginlinemode
   \singlespace \raggedcenter}

\def\affil			
  {\vskip 3pt plus 0.1fill \beginlinemode
   \oneandahalfspace \raggedcenter \sl}

\def\abstract			
  {\vskip 3pt plus 0.3fill \beginparmode
   \doublespace ABSTRACT: }

\def\submit  			
	{\vskip 24pt \beginlinemode
	\noindent \rm Submitted to: \sl}

\def\endtopmatter		
  {\endpage			
   \body}

\def\body			
  {\beginparmode}		

\def\head#1{			
  \goodbreak\vskip 0.5truein	
  {\immediate\write16{#1}
   \raggedcenter \uppercase{#1}\par}
   \nobreak\vskip 0.25truein\nobreak}

\def\beneathrel#1\under#2{\mathrel{\mathop{#2}\limits_{#1}}}

\def\refto#1{$^{#1}$}		

\def\references			
  {\head{References}		
   \beginparmode
   \frenchspacing \parindent=0pt \leftskip=1truecm
   \parskip=8pt plus 3pt \everypar{\hangindent=\parindent}}

\gdef\refis#1{\item{#1.\ }}			

\gdef\journal#1, #2, #3, 1#4#5#6{		
    {\sl #1~}{\bf #2}, #3 (1#4#5#6)}		

\def\pra{\journal Phys. Rev. A, }

\def\prb{\journal Phys. Rev. B, }

\def\prl{\journal Phys. Rev. Lett., }

\def\endreferences{\body}

\def\figurecaptions		
  {\endpage
   \beginparmode
   \head{Figure Captions}
}

\def\endpage			
  {\vfill\eject}

\def\endpaper			
  {\endmode\vfill\supereject}

\def\endit
  {\endpaper\end}


\def\heading				
  {\vskip 0.5truein plus 0.1truein	
   \beginparmode \def\\{\par} \parskip=0pt \singlespace \raggedcenter}

\def\subheading				
  {\vskip 0.25truein plus 0.1truein	
   \beginlinemode \singlespace \parskip=0pt \def\\{\par}\raggedcenter}

\def\tag#1$${\eqno(#1)$$}

\def\align#1$${\eqalign{#1}$$}

\def\aligntag#1$${\gdef\tag##1\\{&(##1)\cr}\eqalignno{#1\\}$$
  \gdef\tag##1$${\eqno(##1)$$}}

\def\endaligntag{}

\def\overset#1\to#2{{\mathop{#2}^{#1}}}
\def\underset#1\to#2{{\mathop{#2}_{#1}}}

\def\enddocument{\endit}


\def\ref#1{Ref.~#1}			
\def\Ref#1{Ref.~#1}			
\def\[#1]{[\cite{#1}]}
\def\cite#1{{#1}}
\def\(#1){(\call{#1})}
\def\call#1{{#1}}
\def\taghead#1{}
\def\frac#1#2{{#1 \over #2}}

\def\12{{1\over2}}

\def\sla{\raise.15ex\hbox{$/$}\kern-.57em}
\def\leaderfill{\leaders\hbox to 1em{\hss.\hss}\hfill}
\def\twiddle{\lower.9ex\rlap{$\kern-.1em\scriptstyle\sim$}}
\def\bigtwiddle{\lower1.ex\rlap{$\sim$}}
\def\gtwid{\mathrel{\raise.3ex\hbox{$>$\kern-.75em\lower1ex\hbox{$\sim$}}}}
\def\ltwid{\mathrel{\raise.3ex\hbox{$<$\kern-.75em\lower1ex\hbox{$\sim$}}}}
\def\square{\kern1pt\vbox{\hrule height 1.2pt\hbox{\vrule width 1.2pt\hskip 3pt
   \vbox{\vskip 6pt}\hskip 3pt\vrule width 0.6pt}\hrule height 0.6pt}\kern1pt}
\def\tdot#1{\mathord{\mathop{#1}\limits^{\kern2pt\ldots}}}

\def\pmb#1{\setbox0=\hbox{#1}%
  \kern-.025em\copy0\kern-\wd0
  \kern  .05em\copy0\kern-\wd0
  \kern-.025em\raise.0433em\box0 }

\catcode`@=11
\newcount\r@fcount \r@fcount=0
\newcount\r@fcurr
\immediate\newwrite\reffile
\newif\ifr@ffile\r@ffilefalse
\def\w@rnwrite#1{\ifr@ffile\immediate\write\reffile{#1}\fi\message{#1}}

\def\writer@f#1>>{}
\def\referencefile{
  \r@ffiletrue\immediate\openout\reffile=\jobname.ref%
  \def\writer@f##1>>{\ifr@ffile\immediate\write\reffile%
    {\noexpand\refis{##1} = \csname r@fnum##1\endcsname = %
     \expandafter\expandafter\expandafter\strip@t\expandafter%
     \meaning\csname r@ftext\csname r@fnum##1\endcsname\endcsname}\fi}%
  \def\strip@t##1>>{}}

\def\citeall#1{\xdef#1##1{#1{\noexpand\cite{##1}}}}
\def\cite#1{\each@rg\citer@nge{#1}}	

\def\each@rg#1#2{{\let\thecsname=#1\expandafter\first@rg#2,\end,}}
\def\first@rg#1,{\thecsname{#1}\apply@rg}	
\def\apply@rg#1,{\ifx\end#1\let\next=\relax
\else,\thecsname{#1}\let\next=\apply@rg\fi\next}

\def\citer@nge#1{\citedor@nge#1-\end-}	
\def\citer@ngeat#1\end-{#1}
\def\citedor@nge#1-#2-{\ifx\end#2\r@featspace#1 
  \else\citel@@p{#1}{#2}\citer@ngeat\fi}	
\def\citel@@p#1#2{\ifnum#1>#2{\errmessage{Reference range #1-#2\space is bad.}
    \errhelp{If you cite a series of references by the notation M-N, then M and
    N must be integers, and N must be greater than or equal to M.}}\else%
 {\count0=#1\count1=#2\advance\count1
by1\relax\expandafter\r@fcite\the\count0,%
  \loop\advance\count0 by1\relax
    \ifnum\count0<\count1,\expandafter\r@fcite\the\count0,%
  \repeat}\fi}

\def\r@featspace#1#2 {\r@fcite#1#2,}	
\def\r@fcite#1,{\ifuncit@d{#1}		
    \expandafter\gdef\csname r@ftext\number\r@fcount\endcsname%
    {\message{Reference #1 to be supplied.}\writer@f#1>>#1 to be supplied.\par
     }\fi%
  \csname r@fnum#1\endcsname}

\def\ifuncit@d#1{\expandafter\ifx\csname r@fnum#1\endcsname\relax%
\global\advance\r@fcount by1%
\expandafter\xdef\csname r@fnum#1\endcsname{\number\r@fcount}}

\let\r@fis=\refis			
\def\refis#1#2#3\par{\ifuncit@d{#1}
    \w@rnwrite{Reference #1=\number\r@fcount\space is not cited up to now.}\fi%
  \expandafter\gdef\csname r@ftext\csname r@fnum#1\endcsname\endcsname%
  {\writer@f#1>>#2#3\par}}

\def\r@ferr{\endreferences\errmessage{I was expecting to see
\noexpand\endreferences before now;  I have inserted it here.}}
\let\r@ferences=\references
\def\references{\r@ferences\def\endmode{\r@ferr\par\endgroup}}

\let\endr@ferences=\endreferences
\def\endreferences{\r@fcurr=0
  {\loop\ifnum\r@fcurr<\r@fcount
    \advance\r@fcurr by 1\relax\expandafter\r@fis\expandafter{\number\r@fcurr}%
    \csname r@ftext\number\r@fcurr\endcsname%
  \repeat}\gdef\r@ferr{}\endr@ferences}


\let\r@fend=\endpaper\gdef\endpaper{\ifr@ffile
\immediate\write16{Cross References written on []\jobname.REF.}\fi\r@fend}

\catcode`@=12

\citeall\refto		
\citeall\ref		%
\citeall\Ref		%

\catcode`@=11
\newcount\tagnumber\tagnumber=0

\immediate\newwrite\eqnfile
\newif\if@qnfile\@qnfilefalse
\def\write@qn#1{}
\def\writenew@qn#1{}
\def\w@rnwrite#1{\write@qn{#1}\message{#1}}
\def\@rrwrite#1{\write@qn{#1}\errmessage{#1}}

\def\taghead#1{\gdef\t@ghead{#1}\global\tagnumber=0}
\def\t@ghead{}

\expandafter\def\csname @qnnum-3\endcsname
  {{\t@ghead\advance\tagnumber by -3\relax\number\tagnumber}}
\expandafter\def\csname @qnnum-2\endcsname
  {{\t@ghead\advance\tagnumber by -2\relax\number\tagnumber}}
\expandafter\def\csname @qnnum-1\endcsname
  {{\t@ghead\advance\tagnumber by -1\relax\number\tagnumber}}
\expandafter\def\csname @qnnum0\endcsname
  {\t@ghead\number\tagnumber}
\expandafter\def\csname @qnnum+1\endcsname
  {{\t@ghead\advance\tagnumber by 1\relax\number\tagnumber}}
\expandafter\def\csname @qnnum+2\endcsname
  {{\t@ghead\advance\tagnumber by 2\relax\number\tagnumber}}
\expandafter\def\csname @qnnum+3\endcsname
  {{\t@ghead\advance\tagnumber by 3\relax\number\tagnumber}}

\def\equationfile{%
  \@qnfiletrue\immediate\openout\eqnfile=\jobname.eqn%
  \def\write@qn##1{\if@qnfile\immediate\write\eqnfile{##1}\fi}
  \def\writenew@qn##1{\if@qnfile\immediate\write\eqnfile
    {\noexpand\tag{##1} = (\t@ghead\number\tagnumber)}\fi}
}

\def\callall#1{\xdef#1##1{#1{\noexpand\call{##1}}}}
\def\call#1{\each@rg\callr@nge{#1}}

\def\each@rg#1#2{{\let\thecsname=#1\expandafter\first@rg#2,\end,}}
\def\first@rg#1,{\thecsname{#1}\apply@rg}
\def\apply@rg#1,{\ifx\end#1\let\next=\relax%
\else,\thecsname{#1}\let\next=\apply@rg\fi\next}

\def\callr@nge#1{\calldor@nge#1-\end-}
\def\callr@ngeat#1\end-{#1}
\def\calldor@nge#1-#2-{\ifx\end#2\@qneatspace#1 %
  \else\calll@@p{#1}{#2}\callr@ngeat\fi}
\def\calll@@p#1#2{\ifnum#1>#2{\@rrwrite{Equation range #1-#2\space is bad.}
\errhelp{If you call a series of equations by the notation M-N, then M and
N must be integers, and N must be greater than or equal to M.}}\else%
 {\count0=#1\count1=#2\advance\count1
by1\relax\expandafter\@qncall\the\count0,%
  \loop\advance\count0 by1\relax%
    \ifnum\count0<\count1,\expandafter\@qncall\the\count0,%
  \repeat}\fi}

\def\@qneatspace#1#2 {\@qncall#1#2,}
\def\@qncall#1,{\ifunc@lled{#1}{\def\next{#1}\ifx\next\empty\else
  \w@rnwrite{Equation number \noexpand\(>>#1<<) has not been defined yet.}
  >>#1<<\fi}\else\csname @qnnum#1\endcsname\fi}

\let\eqnono=\eqno
\def\eqno(#1){\tag#1}
\def\tag#1$${\eqnono(\displayt@g#1 )$$}

\def\aligntag#1\endaligntag
  $${\gdef\tag##1\\{&(##1 )\cr}\eqalignno{#1\\}$$
  \gdef\tag##1$${\eqnono(\displayt@g##1 )$$}}

\def\eqalignno#1{\displ@y \tabskip\centering
  \halign to\displaywidth{\hfil$\displaystyle{##}$\tabskip\z@skip
    &$\displaystyle{{}##}$\hfil\tabskip\centering
    &\llap{$\displayt@gpar##$}\tabskip\z@skip\crcr
    #1\crcr}}

\def\displayt@gpar(#1){(\displayt@g#1 )}

\def\displayt@g#1 {\rm\ifunc@lled{#1}\global\advance\tagnumber by1
        {\def\next{#1}\ifx\next\empty\else\expandafter
        \xdef\csname @qnnum#1\endcsname{\t@ghead\number\tagnumber}\fi}%
  \writenew@qn{#1}\t@ghead\number\tagnumber\else
        {\edef\next{\t@ghead\number\tagnumber}%
        \expandafter\ifx\csname @qnnum#1\endcsname\next\else
        \w@rnwrite{Equation \noexpand\tag{#1} is a duplicate number.}\fi}%
  \csname @qnnum#1\endcsname\fi}

\def\ifunc@lled#1{\expandafter\ifx\csname @qnnum#1\endcsname\relax}

\let\@qnend=\end\gdef\end{\if@qnfile
\immediate\write16{Equation numbers written on []\jobname.EQN.}\fi\@qnend}

\catcode`@=12


\title Hysteresis and hierarchies: dynamics of disorder--driven first--order %
phase transformations
\author  James~P.~Sethna, Karin~Dahmen, Sivan~Kartha,
James~A.~Krumhansl\footnote{$^\dagger$}{Present Address: %
515 Station Rd., Amherst, Massachusetts 01002},%
Bruce~W.~Roberts, and Joel~D.~Shore\footnote{$^*$}{Present %
Address: Department of Physics, Simon Fraser University, %
Burnaby, British Columbia, Canada  V5A 1S6}%
\affil Laboratory of Atomic and Solid State Physics, Cornell University, %
Ithaca, New York 14853-2501

\abstract
We use the zero--temperature random--field Ising model to study hysteretic
behavior at first--order phase transitions.  Sweeping the external field
through zero, the model exhibits hysteresis, the return--point memory effect,
and avalanche fluctuations.  There is a critical value of disorder at which a
jump in the magnetization (corresponding to an infinite avalanche)
first occurs.
We study the universal behavior at this critical point using
mean--field theory, and also present preliminary results
of numerical simulations in three dimensions.

\noindent PACS numbers: 75.60.E, 64.60.Ht, 64.60.My, 81.30.Kf, 05.50.+q

\noindent

\endtopmatter

\noindent
\nobreak

First--order phase transitions have always been the weak sibling of critical
phenomena in statistical mechanics.   At some critical temperature
$T_c$, the idealized thermodynamic equilibrium properties of
a homogeneous material abruptly shift from (say) liquid to gas, with precursor
fluctuations almost entirely absent.
The classic model for a first order
transition\refto{Fisher} is the Ising model in an external field $H$ at
$T<T_c$: as $H$ passes through zero, the equilibrium magnetization
reverses abruptly.

There is an amazing contrast, though, with real first--order transitions as
studied by materials scientists and metallurgists (figure~1).  A
solid material which transforms from one crystalline or magnetic
form to another under the influence of temperature, external stress,
or applied field often
has no sharp transition at all.  Hysteresis becomes the key phenomenon.
Also, there are the experimental Barkhausen\refto{Barkhausen} and
return--point memory\refto{RPME} effects, which are explained here in terms
of avalanches and a hierarchy of metastable states.

The ingredient we will add to the idealized physics picture is
disorder.  Adding
a random field $f_i$ at each site of the Ising model
$${\cal H} = -\sum_{ij} J_{ij}s_i s_j - \sum_i f_i s_i + H s_i
							\eqno(Hamiltonian)$$
allows us to study the effect of disorder in a first order phase
transition. We study this random--field Ising model at zero
temperature. As the external field $H$ is changed,
each spin will flip when the direction of its total local field
$$F_i \equiv \sum_j J_{ij} s_j + f_i + H\eqno(Internal)$$
changes. The system transforms from negative to positive
magnetization as the field is swept upwards.
Related approaches\refto{related, Levy}
have been useful for studying the Barkhausen effect,
but they have not discussed the return--point memory and did not
attempt to vary the disorder and field to reach a
critical point.

The dynamics of the approach to thermal equilibrium in this model has been
well studied\refto{RFIM}. However, here we study the {\it athermal}
dynamical response to an external field, where the transition
comprises a series of cluster flips.
This model is applicable to many experimental systems where the
elementary domains have barriers to flipping large enough that
thermal activation can be ignored. For example,
martensitic transitions come in two varieties, called
isothermal and athermal: the athermal martensites show no transformation if
the temperature and external strain are held constant, but transform
with a crackling noise  as these parameters are changed.
Similarly, most useful magnetic memory devices by design do not come to
thermal equilibrium!

Simulating the system described in equation \(Hamiltonian) at zero
temperature yields behavior pleasantly familiar to the
experimentalist.  Figure~1 shows the
hysteresis loop for a three--dimensional $30^3$ system with
nearest--neighbor bond strength $J=1$ and random fields ${f_i}$ with a
Gaussian probability distribution with a half-width $R$.  The outer loop
shows the external field $H$ being swept from a large negative value
(saturating all spins to -1) to a large positive value and back.

The inner loops in figure~1 illustrate the {\it return--point memory effect},
seen in some but by no means all first order transitions\refto{RPME}.
(Some first--order transitions exhibit a drift in
their hysteretic behavior.\refto{Levy})
If the field $H$ is made to backtrack from $H_B$ to $H_C$, when it returns to
$H_B$ the system returns precisely to the same state from which it left the
outer loop, it {\it remembers} the former state.  The slope of the $M(H)$
curve has a nonanalyticity as it rejoins the outer loop:  when the field
$H(t)$ passes through the previous local maximum $H_B$,
new spins begin to flip, leading to the
apparent slope discontinuity at point {\bf B}.  This same memory effect
extends to subcycles within cycles (and so on); {\it the
state of the system can remember an entire hierarchy of
turning points in its past external field} $H(t)$.

This memory effect is vividly illustrated by experiments measuring the
avalanche fluctuations during the transformation.  The nucleation of the
individual domains in martensites can produce observable pulses in acoustic
emission and latent heat.\refto{Acoustic} Under a cycle like
that between $H_B$ and $H_C$ in figure~1, the acoustic emission is resolved
into hundreds of individual pulses each lasting some $\mu$sec, of varying
height, forming a ``fingerprint'' of the cycle.  On repeating the
cycle, the avalanche structure is precisely reproduced!

We can explain the return--point memory effect theoretically using
Alan Middleton's ``No
Passing'' rule, introduced in the study of sliding charge--density
waves.\refto{Middleton} Consider the natural partial ordering of the states: a
state ${\bf s} = \{s_1, \dots, s_N\} \ge {\bf r} =
\{r_1, \dots, r_N\}$ if $s_i \ge r_i$ for each site $i$ in the system.  This
ordering is not very discriminating: most arbitrary pairs of states
will not have any
definite relationship.  For example, in figure~1, the states {\bf
a} and {\bf b} are unrelated: despite the fact that the net magnetization of
{\bf a} is smaller than that of {\bf b}, there is likely at least one spin
that has flipped on the way from {\bf a} to {\bf B} which has not flipped
back down on the way down to {\bf b}.  This partial ordering
becomes important because it is preserved by the
dynamics:

\noindent{\bf No~Passing:} (Middleton) Let a system ${\bf s}(t)$ be evolved
under the field $H_s(t)$, and similarly ${\bf r}(t)$ evolved under
$H_r(t)$.  Suppose the initial configurations satisfy ${\bf s}(0) \ge {\bf
r}(0)$ and the fields $H_s(t) \ge H_r(t)$.  Then it will remain true that
${\bf s}(t) \ge {\bf r}(t)$ at all times $t$.

\noindent{\bf Proof:}~Suppose the contrary.  Then there must be some first
time that a spin $j$ in {\bf r} is going to ``pass'' the corresponding spin
in {\bf s}, {\it i.e.} $r_j(t) > s_j(t)$.  At that time, the local field
$F^r_j$ must be larger than $F^s_j$.  But this cannot be, because the
neighbors of $r_j$ in state ${\bf r}$ at that time are no less negative than
those of $s_j$ and the external field $H_r \le H_s$.

Having established the No~Passing property of the system, we make the
additional assumption that the system dynamics is adiabatic:
the field changes slowly
enough that if we start in some state {\bf A} any {\it monotonic} path from
field $H_A$ to $H_B$ will cause the state to evolve in the same way, and
into the same final state {\bf B}.  For this model and for other
systems with partial ordering, No~Passing, and adiabaticity, we can now prove:

\noindent{\bf Return--Point Memory:} Suppose a system ${\bf s}(0)$ is evolved
under field $H(t)$, where $H(0) \le H(t) \le H(T)$ for $0<t<T$, with $H(t)$
not necessarily monotonic.  Then the final state of the system depends
only on $H(T)$, and is independent of the time $T$ or the history $H(t)$.
In particular, a system coming back to a previous extremal field will return
to exactly the same state, provided that the field remains within these
bounds.

\noindent{\bf Proof:} Consider the fields $H_{\rm min}(t) =
\min_{t'\ge t} H(t')$ and $H_{\rm max}(t) = \max_{t'\le t} H(t')$.
These fields irregularly but monotonically rise from $H(0)$ to $H(T)$.
If ${\bf s}_{\rm min}$ evolves under $H_{\rm min}$ and similarly
${\bf s}_{\rm max}$ evolves under $H_{\rm max}$, then since
$H_{\rm min}(t) \le H(t) \le H_{\rm max}(t)$ the No Passing rule
implies ${\bf s}_{\rm min}(t) \le {\bf s}(t) \le {\bf s}_{\rm max}(t)$.
But by the adiabaticity assumption, all monotonic paths lead to the same
final state ${\bf s}_{\rm min}(T) = {\bf s}_{\rm max}(T)$, so
${\bf s}(T)$ is independent of path.

Historically, ours is not the first model to exhibit hysteresis
and return--point
memory.  These phenomena are often described by Preisach
models\refto{Preisach}, where the system is decomposed into independent
elementary hysteresis domains, each with an upper and lower critical field
for flipping.  Preisach models have the three properties needed for
return--point memory, and indeed can be thought of as  zero--dimensional
variants of our model.  They are successful because they demonstrate
return--point memory, and because the distribution of domains can be varied
in order to fit experimental hysteresis loops.  The hypothesis of independent
domains, of course, is an idealization, since there is generally some
non-trivial coupling between components of the system.
As an illustration of the collective behavior missing in the
Preisach model, we unearth critical fluctuations and
universality buried in the dynamics of the interacting system.

Physically, if the nearest--neighbor coupling is substantial
compared to the randomness,
{\it i.e.} bond strength $J {\gg} R$, then the system will look
like what one expects of a clean first order transition:  the
first sufficiently large region to nucleate will push most of the rest of
the sample over the brink.  In this small disorder regime, there will be an
``infinite avalanche'':
some spin flips, triggering its neighbors to flip
(and so on) until a finite fraction of the sample is transformed, causing a
jump in the magnetization density.  This infinite avalanche
will perhaps be surrounded by precursors and aftershocks.
On the other hand, in the large disorder regime, where $J {\ll} R$ the spins
will essentially flip on their own, each spin flipping as the external field
crosses its local random field. Those avalanches which occur will be small:
there will be no diverging correlation lengths.

Consider the ``phase diagram'' of our model as we vary disorder $R$ and
field $H$ (at $T=0$), starting from the state with all spins down (figure~2).
There must be a strength of disorder $R_c$ above which an
infinite avalanche never happens.  In contrast,
for weaker disorder, there will be a field $H_c^u(R)$ at which
(on raising the field from the down state) an infinite avalanche occurs.
Both experimentally\refto{burst} and in our initial simulations in
three dimensions,
the transition at $H_c^u(R)$ seems to be abrupt for $R<R_c$. (We'd call it
first--order, except the avalanche is supposed to be mediating a
first--order transition --- the language is failing us.)
On the other hand, as
one approaches the critical field $H_c^u(R_c)$ at the critical disorder
$R_c$, the transition appears to be continuous:  the magnetization $M(H)$
has a power--law singularity, and there are avalanches of all sizes.  As one
approaches this ``endpoint'' at $(R_c, H_c^u(R_c))$ in the $(R,H)$ plane, we
find diverging correlation lengths and presumably universal critical
behavior.

We can solve for these critical properties exactly within mean--field
theory.  Suppose every spin in equation \(Hamiltonian) is coupled to all $N$
other spins with coupling $J/N$.  The effective field acting on a site is $J
M + f_i + H$, where $M = \sum_i s_i/N$ is the average magnetization.  Spins
with $f_i < -J M - H$ will point down, the rest will point up.  Thus the
magnetization $M(H)$ is given implicitly by the expression
$$M(H) = 1 - 2\int_{-\infty}^{-JM(H)-H} \rho(f) \, df \eqno(consistent)$$
where $\rho(f)$
is the probability distribution for the random field.  This equation has a
single--valued solution unless  $R \le R_c$ (which in the case
of a Gaussian distribution corresponds to  $\rho(0) \ge 1/ 2J$),
at which point hysteresis and an infinite avalanche begin.
Near the endpoint, the jump in the magnetization
$\Delta M$ due to the avalanche scales as $r^\beta$, where $r \equiv
(R_c-R)/R_c$ and $\beta=1/2$.  As one varies both $r$ and the reduced field
$h \equiv (H-H_c^u(R_c))$, the magnetization scales as $M(h,r)
\sim |r|^\beta {\cal M}_\pm(h/|r|^{\beta \delta})$, where $\pm$ refers to the
sign of $r$.
In mean--field theory
$\delta = 3$ and ${\cal M}_\pm$ is given by the smallest real root
$g_\pm (y)$ of the cubic equation
$g^3 \mp {12 \over \pi } g-{ 12 \sqrt 2  \over \pi^{3/2} R_c } y = 0$.

Unfortunately, the mean field theory predicts unphysical
behavior in two ways.  First,
there is no hysteresis apart from the infinite avalanche.  This
is an artifact of the Ising mean--field theory, and is not observed
in finite dimensions nor in an otherwise equivalent soft--spin mean--field
model.  Second, the approach to the infinite avalanche upon varying
$H$ for $R < R_c$ is continuous in mean--field theory.
However, in three dimensions we numerically observe a discontinuous
transition, although fluctuation effects do seem to be large.

More interesting is the avalanche size distribution near the critical
point, (inset, figure~2).  We can
solve exactly for the probability $D(s,t)$ of having an
avalanche of size $s$, where ${t\equiv 2J \rho(-JM-H)-1}$
measures the distance to the infinite avalanche at $\rho = 1/2J$.
To have an avalanche of size $s$ triggered by a spin with
random field $f$, you must have precisely $s-1$ spins with random fields in
the range $\{f, f + 2 J s / N \}$.  The probability for this to happen is
given by the Poisson distribution.  In addition, they must be arranged so
that the first spin triggers the rest.  This occurs with probability
precisely $1/s$, which we can see by putting periodic boundary conditions on
the interval $\{f, f + 2 J s / N \}$ and noting that there would be exactly
one spin of the $s$ which will trigger the rest as a single avalanche.  This
leads to the avalanche size distribution $$D(s,t) = {s^{s-2}\over (s-1)!}
(t+1)^{s-1} e^{-s (t+1)}.\eqno(D)$$ To put this in a scaling form, we must
first express $t$ as a function of $r$ and $h$: \break
${t \sim r\left[ 1 \mp {\pi \over 4 } g_\pm (h/|r|^{3 \over 2})^2 \right]}$.
Using some simple expansions and
Stirling's formula, we can then write $D$ in the scaling form
$$D(s,r,h) \sim s^{-\tau}{\cal D}_\pm(s/|r|^{-1/\sigma},
h/|r|^{\beta\delta})\eqno(D2)$$ where our mean--field calculation gives $\tau
= 3/2$, $\sigma = 1/2$, and the universal scaling function
$${\cal D}_\pm(x,y)={1\over\sqrt{2\pi}}e^{-x{\left[ 1 \mp {\pi \over 4 }
g_\pm (y)^2 \right]^2/2}}.\eqno(scaling)$$

As usual, we expect the critical exponents $\beta$, $\tau$, $\delta$, and
$\sigma$ and the scaling functions $\cal M$ and $\cal D$ to be independent
of many details of the system (and thus the same for theoretical models and
real materials), but to depend on dimension, the range of the interaction,
and on the symmetries of the order parameter.
We intend to extract critical exponents both numerically and via
an $\epsilon$-expansion (the upper critical dimension for the closely related
charge--density--wave depinning problem is four \refto{Narayan}), study finite
temperatures, and introduce frustration representing the dipole fields in
magnets and the elastic strain in martensites.

Does the critical behavior predicted here exist in the real world?
An FeNi alloy showed a crossover from athermal to burst (infinite
avalanche) behavior as the
grain size was varied.\refto{burst}  Grain boundaries are not random fields,
but we expect critical fluctuations and scaling where the bursting first
occurs.  The distribution of avalanches in magnetic systems has been
studied, and some preliminary fits to power laws have
been made.\refto{Barkhausen}  Avalanches and
hierarchies have been implicit in the literature since the 1920's; power laws
and critical scaling are the tools needed for understanding the
collective behavior being studied now.

\vfill\break
\bigskip
\centerline{\bf Acknowledgments}

We acknowledge the support of DOE Grant \#DE-FG02-88-ER45364.  JPS would
like to thank the Technical University of Denmark and NORDITA for support
and hospitality.  BWR acknowledges support from the Hertz Foundation.
SK acknowledges support from the Department of Education. We
would like to thank J. Ort\'in, A. Planes, Llouis Ma\~nosa, Teresa Cast\'an,
and Oliver Penrose for helpful conversations.

\bigskip
\vfill\break
\centerline{\bf Figure Captions}

\bigskip
\noindent {\bf Figure~1: Hysteresis Loop Showing Return--Point Memory}.
Shown is the magnetization as a function of external field for a
$30^3$ system with disorder $R=3.5 J$.
Note that the system returns to the original curve at exactly the same
state B that it left, that the returning curve has an apparent slope
discontinuity at B, and that both effects also happen for the internal
subloop.  Thus a state can have a whole hierarchy of parent states
(mothers at increasing fields and fathers at decreasing fields), which
are seen as kinks in the corresponding branch of the $H(M)$ curve.

\bigskip
\noindent {\bf Figure~2: Varying the Disorder}.
Three $H(M)$ curves for different levels of disorder, for a $60^3$ system.
Our current estimate
of the critical disorder is $R_c=2.21 J$ (we set $J=1$).
At $R=2<R_c$, there is an
infinite avalanche which seems to appear abruptly.
For $R=2.6>R_c$ the dynamics is
macroscopically smooth, although of course microscopically it is a sequence
of sizable avalanches.  At $R=2.3$, near the critical level of disorder,
extremely large events become common.
{\bf Inset}: Log-Log Plot of the avalanche-size distribution $D(s)$ vs.
avalanche size $s$
for the $60^3$ system at $R=2.3$ for $1.3<H<1.4$, averaged over 20 systems.
Here $D(s) \sim s^{-1.7}$, compared to the mean--field exponent
$\tau$ of $3/2$.
We expected to see a cutoff in the avalanche sizes: presumably we will
see it farther from the endpoint or for larger system sizes.

\vfill\break

\references

\refis{Fisher} V.~Privman and M.~E. Fisher, {\sl J.~Stat.~Phys.},
{\bf 33}, 385 (1983), M.~E. Fisher and V.~Privman, \prb 32, 447, 1985,
and references therein.

\refis{Preisach} F.~Preisach, {\sl Z.~Phys.} {\bf 94},277 (1935);
M.~Krasnoselskii
and A.~Pokrovskii, {\sl Systems with Hysteresis}, (Nauka, Moscow,1983),
I.~D. Mayergoyz, {\sl J.~Appl. Phys.} {\bf 57}, 3803 (1985);
{\sl Mathematical Models of Hysteresis}, (Springer-Verlag,1991); P.C.
Clapp, {\sl Materials Science and Engineering} {\bf A127}, 189-95 (1990).

\refis{RFIM} J.~Villain, \prl 29, 6389, 1984,
G.~Grinstein and J.~F. Fernandez, \prb 29, 6389, 1984,
A.~J. Bray and M.~A. Moore, {\sl J.\ Phys.\ C} {\bf18}, L927 (1985),
D. S. Fisher, \prl 56, 416, 1986.

\refis{RPME} Return point memory has been seen in ferromagnetism:
D.~C. Jiles, D.~L. Atherton, {\sl J. Appl. Phys.} {\bf 55}, 2115 (1984);
Barker {\it et al.}, {\sl Proc. R. Soc. London A} {\bf 386}, 251-261 (1983).
Martensitic transformations (thermally and stress-induced):
J. Ort\'in {\sl J. Appl. Phys.} {\bf 71}, 1454 (1992); {\sl J.de
Phys. IV}, colloq. C4, {\bf 1}, C-65, (1991) and references therein.
Adsorption of gases: J. Katz, {\sl J.
Phys. (Colloid) Chem.} {\bf 53}, 1166 (1949); Emmett and Cines, {\sl J. Phys.
(Colloid) Chem.}, {\bf 51}, 1248 (1947).  Charge--density waves:
Z.~Z.~Wang and N.~P.~Ong, \prb 34, 5967, 1986.

\refis{Acoustic} A.~Amengual, Ll.~Ma\~nosa, F.~Marco, C.~Picornell,
C.~Segui, V.~Torra, {\sl Thermochimica Acta}, {\bf 116}, 195 (1987);
A.~Planes, J.L.~Macqueron, M.~Morin, G.~Guenin, L.~Delaey, {\sl
J.de Phys.}, colloq. C4, {\bf 12}, C4-615, (1982).

\refis{Middleton} A.~A.~Middleton, \prl 68, 670, 1992; D.~S.~Fisher and
A.~A.~Middleton, preprint.

\refis{Barkhausen}  J. C. McClure, Jr. and K. Schr\"oder, { \it CRC Crit.
Rev. Solid State Sci.} {\bf 6}, 45 (1976);
P. J. Cote and L. V. Meisel, \prl 67, 1334, 1991.

\refis{burst}  V. Raghavan in {\it Martensite}, G.~B.~Olson and
W.~S.~Owen, eds., (ASM International), 1992, p. 197.

\refis{related} G. Bertotti and M. Pasquale, {\it J. Appl. Phys.}
{\bf 67}, 5255 (1990); {\it J. Appl. Phys.} {\bf 69}, 5066 (1991).
J.~V. Andersen and O.~G. Mouritsen, \pra 45, R5331, 1992.

\refis{Narayan}   O.Narayan and D.S.Fisher, \prl 68, 3615, 1992
and preprint (unpublished).

\refis{Levy} L.~P.~L\'evy, ``Reptation and hysteresis in disordered
magnets,'' to be published.

\endreferences

\enddocument

\end